\documentstyle{mn}
\newcommand{\Msun}{\mbox{M$_{\odot}$}}

\newcommand{\ltsimeq}{\raisebox{-0.6ex}{$\,\stackrel 
        {\raisebox{-.2ex}{$\textstyle <$}}{\sim}\,$}} 
\newcommand{\gtsimeq}{\raisebox{-0.6ex}{$\,\stackrel
        {\raisebox{-.2ex}{$\textstyle >$}}{\sim}\,$}}

\newcommand{\chis}{\mbox{$\chi^{2}~$}}

\title[The massive white dwarf in the recurrent nova T CrB]
{The massive white dwarf in the recurrent nova T CrB}
\author[Shahbaz et al.]{
T. Shahbaz,$^{1}$ M. Somers,$^{2}$ B. Yudin,$^{3}$ and T. Naylor$^{2}$ \\
$^{1}$Department of Astrophysics, Nuclear Physics Building, Keble Road,
Oxford, OX1 3RH, UK \\
$^{2}$Department of Physics, Keele University, Keele, Staffordshire, 
ST5 5BG, UK \\
$^{3}$Sternberg Astronomical Institute, University of Moscow, Russia }

\begin{document}

\maketitle

\begin{abstract}

\noindent
We have obtained $I$, $J$, $H$ and $K$ band light curves of the recurrent
nova T CrB. We find that we can only fit the $J$ band light curve with a
Roche-lobe filling secondary star and a dark spot of radius
11$^\circ$--26$^\circ$ (90 per cent confidence) centred at the inner
Lagrangian point. We obtain limits to the binary inclination of
38$^\circ$--46$^\circ$ (90 per cent confidence) which when combined with
the value for the mass function allows us to determine the mass of the
compact object to be 1.3--2.5 \Msun (90 per cent confidence). This mass
range is consistent with the Chandrasekhar limiting white dwarf mass, and
so we provide evidence needed to support the outbursts in recurrent novae
in terms of a thermonuclear runaway process on the surface of a massive
$\sim$1.4 \Msun white dwarf.

\end{abstract}

\section{Introduction}

T CrB is a well studied recurrent nova with an M giant component. It
underwent nova like outbursts in 1866 and 1946, with light curves which
were very similar. They are characterised by having a fast rise to
maximum and by a secondary, fainter maximum occurring $\simeq 100$ days
later. The secondary maximum has a very slow decay time ($>$ 10 years).

Two competing models have been suggested to account for the peculiar
outburst behaviour. One suggestion is that the outburst is due to a
thermonuclear runaway (TNR) on the surface of a white dwarf primary.  If
this is the case then the short recurrence time implies that the white
dwarf must have a mass very close to the Chandrasekhar limit (1.4 \Msun;
Chandrasekhar 1939) and that the ejected envelope should be relatively
small (Webbink 1976a). Alternatively Plavec, Ulrich \& Polidan (1973) and
Webbink (1976a) proposed that the hot component is a normal main sequence
star, with eruptions being caused by sudden mass accretion events. This
model can nicely account for a fading of the system before the outburst
and the double maximum in the outburst light curve, however one is
confined to a main sequence accretor. Selvelli, Cassatella \& Gilmozzi
(1992) have found evidence from IUE spectra that the accretor is a
white dwarf. They point to three pieces of evidence: (1) the bulk of the
luminosity is emitted in the UV with little contribution from the hot
component in the optical, (2) strong $\rm He~\sc II$ and $\rm N~\sc V$
emission lines suggest temperatures of $\simeq 10^{5}K$ and (3) the
observed large rotational broadening of the high excitation lines.

Sandford (1949) first detected radial velocity variations in the emission
lines while Kraft (1958) established a spectroscopic period of the system
to be 227.6 days. Kraft (1958) also derived a mass ratio
($q=M_{1}/M_{2}$) of 0.71 (where $M_{1}$ is the giant component) based in
part on radial velocities extracted from emission lines originating in
material accreting onto the primary.  We now view radial velocities
derived from such emission lines as unsound and thus reject his value for
$q$. Kenyon \& Garcia (1986) obtained a lower limit to the binary mass
ratio of 2.5 by determining an upper limit to the rotational broadening
of the giant star.

In this paper we present a photometric study of the ellipsoidal
variations of the M giant. By fitting the $J$ band light curve we
determine the binary inclination and hence determine the mass of the
compact component.

\begin{table}
\caption{Log of observations}
\begin{center}
\begin{tabular}{lcc} 
Band  & Date of observation  & Number of points \\ \\
 I & Sep 1994 -- Jun 1995 & 420 points \\
 J & Aug 1987 -- Jun 1995 & 102 points \\
 H & Sep 1987 -- Mar 1990 & 12  points \\
 K & Sep 1988 -- Jun 1995 & 58  points \\ \\
\end{tabular}
\end{center}
\end{table}	

\section{Photometric observations}

The infrared observations were taken at the Crimean Astrophysical
Observatory of Sternberg Astronomical Institute using an InSb photometer
on the 1.25m telescope. We obtained $J$ (12500\AA), $H$ (16500\AA) and
$K$ (22000\AA) band photometry of T CrB during the period 1987--1995 (see
Table 1). Most of the data have an accuracy of better than 0.02 mags. The
standard BS 5947 was used as a comparison star and BS 5972 as a check
star.

The $I$ (8500\AA) band data were taken through a Johnson filter with the
0.6m Thornton Reflector at Keele Observatory during 1994--1995 using a
Santa Barbara Instruments ST-6 CCD camera mounted at the f/4.5 Newtonian
focus. The dark current was subtracted for each image using a median
stack of several dark frames. Relative photometry was carried out using
the routine described in Shahbaz, Naylor \& Charles (1994). 

The individual data for each band were then folded on the ephemeris given
by Kenyon \& Garcia (1986) and then binned in orbital phase (see Fig. 1).

\section{The M giant ellipsoidal variations}

The observed ellipsoidal variations are primarily due to the giant star
presenting differing aspects of its distortion as it orbits the compact
object. By measuring the amplitude of this modulation, the binary
inclination can be determined (see Shahbaz, Naylor \& Charles 1993 and
references within). The $J$ band data covers $\sim$11 orbital cycles 
and the individual points fit the same light curve. This suggests
that the M giant does not vary its intrinsic brightness by more than a few
hundredths of a magnitude (Yudin \& Munari 1993). The erratic and
quasi-periodic variations, with amplitudes larger than the ellipsoidal
modulation of the M giant, make interpreting the optical light curves
very difficult (Webbink 1976b; Lines et al. 1988; Peel 1990). These
effects reduce at longer wavelengths, and they should be absent in the
infrared, where the emission from the M giant star dominates. The
amplitude of our $I$ band light curve is $\sim$ 0.35 mags, which is more
than a factor 2 greater than that expected for the ellipsoidal variations of
the M giant inclined at 40$^\circ$ (see section 4). In light of the
quasi-periodic behaviour of the optical light curves, we only fit the $J$
band light curve, where one expects these uncertainties to be reduced.

\section{The classical model}

In the M star of a cataclysmic variable secondary, the envelope is
deeply convective. Sarna (1989) obtained the gravity-darkening using a
modified form of Lucy's (Lucy 1967) gravity-darkening law appropriate
for a star with a convective envelope. One can model the ellipsoidal
variations as a function of the binary mass ratio $q$, inclination
$i$, the  effective temperature of the secondary star T$_{\rm eff}$,
the limb-darkening coefficient and gravity-darkening exponent $\beta$. 

Using the ellipsoidal model described in Shahbaz, Naylor \& Charles
(1993) we performed a least-squares fit to the $J$ band light curves,
grid searching the variables $q$ and $i$. We used T$_{\rm eff}$=3500 K,
$\beta$=0.08 and $i$ in the range 20$^\circ$--90$^\circ$. From optical
spectroscopy, Kenyon \& Garcia (1986) estimate that the mass ratio must
be at least 2.5. We therefore search $q$ in the range 2.0--10.0. The
appropriate limb-darkening coefficient for each wavelength and
temperature was extrapolated using the values given by Al-Naimiy (1978).
We obtained a minimum \chis of $\chi_{\rm \nu}^{2}$=2.9 at $q$=3,
$i$=48$^\circ$. As one can see, this fit (dotted line in Fig. 2) does not
describe the data well.

In an attempt to explain the large difference between the minima we
explored fits using different values for the gravity darkening exponent.
We computed fits by grid searching the variables $q$, $i$ and $\beta$. As
explained in the Appendix, we find that the blackbody assumption for the
$H$ and $K$ band data is poor. Therefore, in what follows we only use the
$J$ band data where the blackbody assumption is more robust. Fig. 3 shows
the \chis fit in the $\beta-i$ plane, obtained by collapsing the minimum 
\chis solutions along the $q$ axis onto the $\beta-i$ plane. In effect we
have let $q$ run as a free parameter. We obtained a minimum \chis of 2.5
at $q$=3, $i$=43$^\circ$ and $\beta$=0.20. The limits derived are
$i$=35--51$^\circ$ and $\beta$=0.05--0.36 (68 per cent confidence). The
68 and 90 per cent confidence regions (solid and dashed lines
respectively) are shown, calculated according to Lampton, Margon \&
Bowyer (1976) for 2 parameters, after the error bars had been scaled to
give a minimum $\chi^{2}_{\nu}$ of 1. The solid line in Fig. 2 shows the
best fit to the $J$ band light curve. As one can see the fit to the data
points near phase 0.5 is not very good. There still is an extra
component which is introducing an uncertainty of about 60 per cent to the
light curves. In the next section we try to explain the large difference
between the minima in the $J$ band light curves in terms of a dark spot
located at the inner Lagrangian point.

\section{The need for a dark spot}

Star-spots have been observed in other interacting binary stars, in
particular in those systems in which the mass-losing component is a
late-type star, e.g. RS CVn systems (Rodono 1983). Also there is much
stellar activity associated with M stars. Peel (1990) presents
marginal evidence for stellar activity in T CrB near phase 0.5 from the
observations of ultraviolet flares (Ianna 1964) and the visual 
brightening of T CrB.

In an attempt to explain the large difference between the minima of the
$J$ band light curve, we fit the light curve with the ellipsoidal
modulation of the secondary star plus a dark spot. Since the maxima in
the light curve are almost equal, for simplicity we centred the spot
around the inner Lagrangian point (L$_{1}$). The spot is described as a
circle extending to a latitude R$_{\rm spot}$$^\circ$ away from the
L$_{1}$ point. The effective temperature in the dark spot region
is lower than it would have been in the absence of the spot by 750 K. This is
the typical observed temperature difference in RS CVn systems and T Tauri
stars (Rodono 1986).

Using the nominal value for the gravity darkening exponent for a
convective star (0.08), we obtained fits in the $q$, $i$ and R$_{spot}$
plane. We find $\chi^{2}_{\rm \nu,min}$ = 2.6 at $i$=40$^\circ$ and
R$_{\rm spot}$=20$^\circ$. The 90 per cent confidence limits are
$i$=38--46$^\circ$ and R$_{\rm spot}$=11--26$^\circ$ (see Fig. 4).
Lowering the spot temperature by a further 250 K changes the 
inclination by $<1^{\circ}$.
The dashed line in Fig. 2 shows the best fits to the $J$ band light curve
using a dark spot of size 20$^\circ$. As one can see the fit to the
light curve has improved. 

That using large values for $\beta$ gives fits of similar
quality to that using a dark spot placed around the L$_{1}$ point is not
surprising. If we look at the temperature distribution across the
secondary star for $\beta$=0.08 and 0.20 (see Fig. 5), then one can see
that the effect of high values for $\beta$ is to simulate two regions of
low temperature around the inner Lagrangian point and the opposite
hemisphere of the secondary star. This effect can to some degree be
reproduced by adding a dark spot round the L$_{1}$ point.

\begin{table*}
\caption{Extreme Evolutionary solutions for T CrB}
\begin{tabular}{lcc} 
Parameter & Minimum mass & Maximum mass \\
\\
Secondary core mass $\rm M_{c}$ ($\rm M_{\odot}$)    & 0.2894  & 0.3401 \\
Secondary total mass $\rm M_{2}$ ($\rm M_{\odot}$)   & 0.2894  & 2.000 \\
Secondary radius $\rm R_{2}$ ($\rm R_{\odot}$)       & 26.0  & 49.44  \\
Secondary luminosity $\rm L_{2}$ ($\rm L_{\odot}$)   & 117.2 & 308.2 \\
Secondary effective temperature $\rm T_{eff}$ ($\rm K$)   & 3731  & 3445 \\
Mass transfer rate $\rm M_{\odot}yr^{-1}$          & 3.8 $\times 10^{-9}$ & 
5.8$\times 10^{-8}$  \\
Primary mass $\rm M_{1}$ ($\rm M_{\odot}$)         & $>$ 0.64 & $>$ 1.56 \\
\end{tabular}
\end{table*}	

\section{The evolutionary status of the secondary}

In this section we use theoretical equations which describe the
luminosity and radius of the giant star in order to determine the
predicted mass transfer rate. For evolved secondary stars, i.e. stars on
the first red giant branch and asymptotic giant-branch stars Joss,
Rappaport \& Lewis (1987) give the results of fitting the luminosity (L)
and radius (R) core mass (M$_{c}$) relations to numerical models, for
0.17 \Msun $\ltsimeq M_{c} \ltsimeq $1.4 \Msun. They give the
parameterisations

\begin{equation}
\frac{L_{2}}{L_{\odot}} = \frac{ 10^{5.3} M_{c}^{6} }
{ 1 + 10^{0.4} M_{c}^{4} + 10^{0.5} M_{c}^{5} }
\end{equation}

\begin{equation}
\frac{R_{2}}{R_{\odot}} = \frac{ 3.7 \times 10^{3} M_{c}^{6} }
{ 1 + M_{c}^{3} + 1.75 M_{c}^{4} }.
\end{equation}

As noted by King (1988), there is little radius expansion for M$_{c}
\gtsimeq$ 0.7 and so the mass transfer rates will be very low. However,
for M$_{c} \ltsimeq$ 0.7 the denominators in the above equations are
$\sim$ 1. So if mass transfer is to occur, the stellar radius R$_{2}$
must equal the Roche-lobe radius given by Paczynski's formula Paczynski
(1971)

\begin{equation}
\frac{R_{L}}{a} 
= 0.462 \left( \frac{ M_{2} }{ M_{1} + M_{2} } \right)^{1/3}.
\end{equation}

\noindent
Eliminating the binary separation $a$ by use of Kepler's law with the
binary period of 227.53 days, the requirement R$_{2}$=R$_{L}$ implies

\begin{equation}
0.0011 = M_{c}^{6} M_{2}^{-0.5}.
\end{equation}

The limiting cases for M$_{c}$ are given by the extreme values of
M$_{2}$. We must have M$_{2}\ge$M$_{c}$, while the secondary would have
not left the main sequence if M$_{c}$/M$_{2}$ were less than the
Sch\"onberg-Chandrasekhar limiting value of 0.17, implying M$_{2} \le$
5.88 M$_{c}$.

Following King (1993) we find the following values for the two limiting
cases (see Table 2). The limiting cases give M$_{c}$, which when used in
equations (1) and (2) give L$_{2}$ and R$_{2}$. Using
Stefan's Law then gives T$_{\rm eff}$. King (1988) gives the mass transfer
rate as a function of M$_{c}$ and $M_{2}$

\begin{equation}
-\dot{M_{2}} = 6.4 \times 10^{-6} M_{2} M_{c}^{5} \;\;\; {\rm M_{\odot}yr^{-1}}
\end{equation}

\noindent
which can also be used to obtain the predicted mass transfer rates for
the two limiting cases. The lower limit to the compact object mass can
also be determined by using the value of the mass function and
assuming maximum inclination and the two limiting values for M$_{2}$. The
maximum and minimum solutions correspond to T CrB being near the
beginning and end of its evolution respectively. One can see that the
observed spectral type for the secondary star, M3 III (T$_{\rm eff}$=
3500 K) and the observed average mass accretion rate of 2.3$\times
10^{-8}$ \Msun~yr$^{-1}$ (Selvelli et al.~1992) lie in between the
predicted solutions.

\section{Discussion}

\subsection{The mass of the white dwarf} 

Here we determine limits to the mass of the compact object using the
limits derived for the binary inclination by fitting the $J$ band light
curve. Using the equation for the mass function (f(M)=0.30$\pm$0.01
\Msun; Kenyon \& Garcia 1986), with values for $i$ and M$_{2}$, we
can determine M$_{1}$. Fig. 6 shows the M$_{1}$, M$_{2}$ solutions at the
68 and 90 per cent confidence levels. We also show the limits placed on
the mass ratio of $>$ 2.5 and the evolutionary constraints on the mass of
the secondary star (see section 6). It can be seen that the mass of the
compact object is forced to lie in the range 1.3--2.5 \Msun (90 per cent
confidence).

Ramseyer et al.(1993) and Harrison, Johnson \& Spyromilio (1993) show low
resolution $K$ band spectra of T CrB. They find that the spectra
resemble that of a late type giant star, and comparison with M giants
from Kelinmann \& Hall (1986) suggests a spectral type of M2--M5 III.
Fig. 7 shows the flux distribution of T CrB. The optical and infrared
magnitudes were taken from Munari (1992) and were dereddened using
E$_{B-V}$=0.15 (Selvelli, Cassatella \& Gilmozzi 1992). As one see the
flux distribution of T CrB can be well described by a M3 III Kurucz model
atmosphere spectrum (Kurucz 1992). Both the infrared spectra and the flux
distribution of T CrB suggest that there is little contamination of the
IR flux. A 10 per cent contamination of the $J$ band flux would increase
the binary inclination by $\sim$3$^\circ$, which would decrease the 
mass of the compact object by only $\sim$0.3 \Msun.

\subsection{The nature of the white dwarf}

The main ingredients of the thermonuclear runaway (TNR) models proposed
to explain the outburst behaviour in recurrent novae are 

\begin{itemize}
\item The luminosity at maximum light needs to be greater than the Eddington
luminosity, 
\item the mass accretion rates during quiescence must be high, and
\item the white dwarf must be very massive ($\sim$1.4 \Msun).
\end{itemize}

\noindent
Selvelli, Castella \& Gilmozzi (1992) have shown that the outburst in T
CrB was indeed super Eddington and that the mass accretion rate during
quiescence is very high (2.3$\times 10^{-8}$ \Msun~yr$^{-1}$). In the
TNR model the white dwarf mass needs to be close to the Chandrasekhar
maximum mass for a white dwarf of 1.4 \Msun, or else it cannot accrete
enough material from the red giant companion in a short enough time to
produce the short recurrence time.

We have obtained limits to the mass of the compact object to be 1.3--2.5
\Msun, which is  consistent with the Chandrasekhar maximum mass for a
white dwarf of 1.4 \Msun (Chandrasekhar 1939). In Fig. 8 we show for
comparison the mass distribution of white dwarfs, taken from Ritter \&
Kolb (1995). The compact object cannot be a neutron star because of the
observed nova explosions; in neutron stars the strong gravitational
potential prevents material being ejected. Our mass estimates for the
compact object are such that we can support a massive 1.4 \Msun white
dwarf thus providing support for the presently untested predictions of
TNR theory; namely that recurrent novae occur on massive white dwarfs.
White dwarfs with masses $>$ 1.2 \Msun ~and mass accretion rates
$>10^{-8}$ \Msun~yr$^{-1}$ are expected to be net accretors (Livio \&
Turan 1992). Since white dwarfs are not generally born with masses of 1.4
\Msun, our mass determination implies that the white dwarf is a net
accretor; eventually it will exceed the Chandrasekhar critical mass and
become a type Ia supernova.

\section{Conclusions}

We have tried to fit the observed $J$ band light curve of T CrB with the
classical Roche-lobe filling secondary star model with a gravity
darkening exponent of 0.08. However, in order to explain the large
difference between the minima in the $J$ band light curve, we find that
we can only fit the light curve with the addition of a dark spot centred
at the inner Lagrangian point. We obtain the 90 per cent confidence limits
to the binary inclination of 38$^\circ$--46$^\circ$, which when combined
with the mass function gives the mass range for the compact object to be
1.3--2.5 \Msun (90 per cent confidence). This mass range is consistent
with Chandrasekhar limiting white dwarf mass $\sim$1.4 \Msun.

\section*{Acknowledgements}

We are grateful to those who support Keele Observatory, especially the
local amateur astronomers who help maintain the telescopes, and ICL who
have provided computing facilities. TS was supported by a PPARC
postdoctoral fellowship, and TN by a PPARC Advanced fellowship.
BY was supported by RFFI grants 96-02-16353 and 96-02-16794.
The data analysis was carried out on the Keele and Oxford Starlink nodes
using the $\sc ark$ software.

\section*{Appendix}

In our model for the ellipsoidal variations of T CrB, we have assumed
that each element on the surface of the secondary star emits blackbody
radiation of a given temperature.  The temperature of each element is
governed by the gravity darkening law that is adopted, depending on the
structure of the envelope of the secondary star. A real M giant will
obviously not emit as a blackbody, but have rather a complex spectrum
with many late type absorption features. The flux emanating per unit
surface area of such an atmosphere may be significantly different to that
from the same temperature blackbody. We can estimate the accuracy of the
blackbody assumption for observations taken through different passbands
by comparing the ratio of the blackbody fluxes to the model atmosphere
fluxes for different temperatures.

The amplitude of an ellipsoidal light curve can to some degree be
approximated by determining the ratio of flux emitted at two different
temperatures, governed by the mean temperature of the secondary star at
phase 0.0 and 0.5. In order to estimate this temperature change we
computed the ellipsoidal models with no surface temperature variation
($\beta$=0.0) and the maximum expected variation ($\beta$=0.25; see von
Zeipel 1924). We then calculate the mean temperature difference between
the gravity darkened inner face of the secondary and the unaltered
uniform inner face of the secondary and find that the inner face of
the secondary is on average cooler than the rest of the star by about 400
K.

We estimate the accuracy of the blackbody assumption by determining the
magnitude difference between blackbody spectra at 3500 and 3100 K
(after they have been folded through the response of the $J$, $H$ and $K$
filters) and comparing these values with those obtained using model
atmospheres. The model atmosphere spectra for $\log$ g = 4.0 were taken
from Allard \& Hauschildt (1995). We find the departure from the blackbody
assumption in the $J$, $H$ and $K$ bands to be about 2, 9, and 24 per
cent respectively.

\begin{table}
\caption{Comparison of blackbody and model atmospheres fluxes}
\begin{center}
\begin{tabular}{lcc}
	 		  & Blackbody &  Model atmospheres \\
                          &  (mags)   &  (mags)            \\ \\   
% I$_{3100 K}$-I$_{3500 K}$ &  0.72     &   0.73 \\
J$_{3100 K}$-J$_{3500 K}$ &  0.48     &   0.46 \\ 
H$_{3100 K}$-H$_{3500 K}$ &  0.38     &   0.47 \\
K$_{3100 K}$-K$_{3500 K}$ &  0.31     &   0.54 \\ 
\end{tabular}
\end{center}
\end{table}	

\end{document}

{\noindent \large \bf Figure captions.} \\ \\

\noindent
{\bf Figure 1.} The individual $I$, $J$, $H$ and $K$ band light curves of
T CrB folded on the orbital ephemeris given by Kenyon \& Garcia (1986).
The data are plotted over 1.5 cycles. \\

\noindent
{\bf Figure 2.} The $J$ band light curve with model fits. The dotted line
is a fit using $q$=3, $\beta$=0.08, $i$=48$^\circ$ and the solid line is
a fit using $q$=3, $\beta$=0.20 and $i$=43$^\circ$. Note that these two
fits are very similar. The dashed line line is a fit using $q$=3,
$\beta$=0.08, $i$=43$^\circ$ and R$_{\rm spot}$=20$^\circ$. As one can
see, the best fit to the light curve is the one with the dark spot (see
text). \\

\noindent {\bf Figure 3.} The 68 percent confidence level solutions for model
fits to the $J$ band data in the $\beta-i$ plane, obtained by collapsing the
minimum \chis solutions along the $q$ axis. The cross marks the best 
solution. \\

\noindent
{\bf Figure 4.} The 68 percent confidence level solutions for model fits
to the $J$ band data using $\beta$=0.08. The solutions in the R$_{\rm
spot}-i$ plane were obtained by collapsing the solutions along the $q$
axis onto the R$_{\rm spot}-i$ plane. The cross marks the best solution. \\

\noindent
{\bf Figure 5.} The temperature distribution of the secondary star
along the line joining the centre of mass of the binary components 
($x/a$; where $x$ is the distance and $a$ is the binary separation).
The solid lines show the distribution using $\beta$=0.08
and 0.20, and the dotted lines show the distribution using $\beta$=0.08
plus a dark spot of radius 20$^\circ$. \\

\noindent
{\bf Figure 6.} The allowed mass range for the compact object obtained by
combining the inclination limits derived from the $J$ band data with the
mass function. The solid and dashed lines are the 68 and 90 per cent
confidence limits respectively. We also show the lower limit to the mass
ratio $q$=2.5 (dotted line) obtained by Kenyon \& Garcia (1986) and the
bounds to the mass of the secondary star (long dashed lines; see section
6). We limit the compact object to lie in the mass range 1.3--2.5
M$_{\odot}$ (90 percent confidence). \\

\noindent
{\bf Figure 7.} The optical and infrared magnitudes of T CrB, taken from
Munari et al. (1992) and dereddened using E$_{B-V}$=0.15 (Selvelli,
Cassatella \& Gilmozzi 1992). The thin weighted line shows a fit with a
M3 III model atmosphere flux distribution (Kurucz 1992). \\

\noindent
{\bf Figure 8.} The mass distribution of white dwarfs. The data were
taken from Ritter \& Kolb (1995). The dashed lines at 0.64 M$_{\odot}$
and 1.4 M$_{\odot}$ mark the observed white dwarf mass (weighted average)
and the Chandrasekhar limiting mass respectively. Also shown is the mass
of the white dwarf in T CrB (90 per cent confidence).